\documentclass{ifacconf}
\pdfoutput=1

\usepackage{graphicx}      % include this line if your document contains figures
\usepackage{natbib}        % required for bibliography
\usepackage{todonotes}
\makeatletter
\newcommand{\pushright}[1]{\ifmeasuring@#1\else\omit\hfill$\displaystyle#1$\fi\ignorespaces}
\newcommand{\pushleft}[1]{\ifmeasuring@#1\else\omit$\displaystyle#1$\hfill\fi\ignorespaces}
\makeatother

\usepackage{amsmath}
\usepackage{amssymb}
\usepackage{tikz,pgfplots}
\usetikzlibrary{backgrounds, intersections, spy}
\usepackage{diagbox}
\usepackage{algorithm}
\usepackage{etoolbox}

\let\classAND\AND
\let\AND\relax
\usepackage{algorithmic}
\let\AND\classAND
\AtBeginEnvironment{algorithmic}{\let\AND\algoAND}

\newcommand{\T}{\mathsf{T}}

\newcommand{\R}{\mathbb{R}}
\newcommand{\N}{\mathbb{N}}
\newcommand{\C}{\mathbb{C}}
\newcommand{\ri}{\mathrm{i}}
\newcommand{\hinf}{{\mathcal{H}_\infty}}

\DeclareMathOperator{\Real}{Re}

\DeclareMathOperator{\argmin}{argmin}

\DeclareMathOperator{\vtf}{vtf}

\DeclareMathOperator{\vtu}{vtu}

\DeclareMathOperator{\vtsu}{vtsu}

\newcommand{\loss}{\mathsf{L}}

\newcommand{\nx}{n}
\newcommand{\enu}{m}
\newcommand{\ny}{m}
\newcommand{\pHsys}{{\Sigma_{\mathrm{pH}}}}

\newcommand{\pHtf}{{H_{\mathrm{pH}}}}

\newcommand{\pJ}{J}
\newcommand{\pR}{R}
\newcommand{\pQ}{Q}
\newcommand{\pB}{B}

\newcommand{\cR}{\mathcal{R}}
\newcommand{\cH}{\mathcal{H}}

\begin{document}
\begin{frontmatter}

\title{Adaptive Sampling for Structure Preserving Model Order Reduction of Port-Hamiltonian Systems} 
% Title, preferably not more than 10 words.

\thanks[footnoteinfo]{This work is supported by the German Research Foundation (DFG) within the project VO2243/2-1: ``Interpolationsbasierte numerische Algorithmen in der robusten Regelung" and the DFG Cluster of Excellence MATH+ within the project AA4-5: ``Energy-based modeling, simulation, and optimization of power systems under uncertainty".}

\author[First,Second]{Paul Schwerdtner} 
\author[First]{Matthias Voigt} 

\address[First]{Technische Universit\"at Berlin, Institut f\"ur Mathematik, Stra{\ss}e des 17. Juni 136, 10623 Berlin, Germany.}
\address[Second]{Corresponding author: schwerdt@math.tu-berlin.de}

\begin{abstract}                % Abstract of not more than 250 words.
  We present an adaptive sampling strategy for the optimization-based structure preserving model order reduction (MOR) algorithm developed in [Schwerdtner, P. and Voigt, M. (2020). Structure preserving model order reduction by parameter optimization, Preprint arXiv:2011.07567]. This strategy reduces the computational demand and the required a priori knowledge about the given full order model, while at the same time retaining a high accuracy compared to other structure preserving but also unstructured MOR algorithms.

  A numerical study with a port-Hamiltonian benchmark system demonstrates the effectiveness of our method combined with its new adaptive sampling strategy. We also investigate the distribution of the sample points.
\end{abstract}


\end{frontmatter}
%===============================================================================

\section{Introduction}

We consider structure preserving model order reduction (MOR) for linear time-invariant port-Hamiltonian (pH) systems of the form
\begin{align}
  \label{eq:pHdefinition}
  \pHsys: \begin{cases}
    \dot x(t) = (\pJ-\pR)\pQ x(t)+\pB u(t), \\
    y(t) = \pB^\T \pQ u(t),
  \end{cases}
\end{align}
where $\pJ, \,\pR, \,\pQ \in \R^{\nx \times \nx}$ and $\pB \in \R^{\nx \times  \enu}$ with $J=-J^\T$, $R\ge 0,$ and $Q\ge 0$. We call $x : \R  \rightarrow \R^{\nx}, u: \R  \rightarrow \R^{\enu},$ and $y: \R  \rightarrow \R^{\ny}$ the \emph{state}, \emph{input}, and \emph{output} of the system, respectively. The \emph{Hamiltonian} $\mathcal{H}: \R^{\nx}  \rightarrow \overline{\R^{+}}:=\{a \in \R: a\ge 0\}$ of $\pHsys$, which describes the total internal energy of the system, is given by
\begin{align*}
  \mathcal{H}(x(t)) = \frac{1}{2}x(t)^\T Q x(t).
\end{align*}
The \emph{transfer function} of $\pHsys$ 
%$\pHtf : \C \supset \Omega  \rightarrow \C^{\enu \times \enu}$ of \eqref{eq:pHdefinition} is obtained by applying the Laplace transform to both equations in \eqref{eq:pHdefinition} and solving for the transformed state. It
is given by
\begin{align*}
  \pHtf(s) := \pB^\T \pQ\left(sI_{\nx}-(\pJ-\pR)\pQ\right)^{-1}\pB.
\end{align*}
It is a real-rational matrix-valued function that describes the input-to-output mapping in the frequency domain.
%In this way, the transfer function describes the frequency-wise input-to-output mapping of a linear dynamical system. 

The use of pH models in systems and control applications comes with several benefits. First, pH models provide a systematic way to model the interaction of different subsystems, particularly across different physical domains or time scales (\cite{Schaft2014}). Moreover, pH systems are automatically \emph{passive} (\cite{Beattie2019}). This further simplifies the use of pH subsystems in a networked system since the power-conserving interconnection of passive systems is again passive and the passivity of a system implies its stability. Furthermore, the pH paradigm allows for structure preserving spatial semi discretizations of partial differential equations (\cite{Serhani2019}) such that systems as defined in \eqref{eq:pHdefinition} can be obtained from first principle modeling.

However, the increased demand for highly accurate models leads to systems with a large state dimension $\nx$ in most practical applications. This often renders the analysis, simulation, or control of such systems computationally demanding if not prohibitively expensive. For that reason, MOR can be employed to compute a surrogate \emph{reduced order model} (ROM) with state dimension $r$ with an input-to-output mapping similar to the \emph{full order model} (FOM) such that $r \ll \nx$. For a comprehensive introduction to MOR we refer the reader to \cite{Ant05}.

Well-established MOR methods such as \emph{balanced truncation} (BT) or the \emph{iterative rational Krylov algorithm} (IRKA) focus on computing ROMs with a high accuracy in terms of the input-to-output mapping but neglect the preservation of additional structural properties of the FOM. In particular, the ROM computed for a given pH model is not necessarily pH. However, when the pH property of that model is used for the automatic construction of a networked system, then the preservation of the pH property is essential. For that reason, the preservation of additional structural properties during MOR is also considered in the literature. 

The method in \cite{Gugercin2012} adapts IRKA to also preserve the pH property of a dynamical system, while the method in \cite{Polyuga2008} (pH-BT) uses a balancing based approach. In our previous study (\cite{SchV2020}), we observe that for both methods the accuracy is drastically reduced compared to non-structure preserving IRKA or BT (this can also be observed in Fig.~\ref{fig:hinferrors}), and develop a new optimization based MOR framework that also preserves the pH structure but has a similar accuracy as the non-structure preserving MOR methods. We also report that recently \cite{BreitenU2021} have developed a structure preserving MOR method built on the \emph{spectral factorization} of a pH system that achieves an accuracy similar to the original IRKA.

The purpose of this work is to overcome one of the major issues with our direct optimization approach as introduced in \cite{SchV2020}. That approach is based on approximating the transfer function of the given FOM at a set of $n_S \in \N$ sample points $S:=\{s_i\}_{i=1\dots n_S} \subset \ri \R$, where $\ri \R$ denotes the imaginary axis. The major issue is that these sample points are fixed and required as user input. However, the approximation quality of the resulting ROM heavily depends on these sample points. Furthermore, the selection of these sample points is not a trivial task. On the contrary, it requires knowledge about the FOM, which is typically computationally expensive to obtain, and an understanding of the methodology that is used to approximate the FOM. Therefore, in this article, we introduce an adaptive sampling strategy, which automatically chooses new sample points as our method progresses. This strategy is based on an accuracy argument introduced in \cite{Apk2018}.

% In our previous numerical experiments we have therefore selected a large number of logarithmically spaced sampling points in the frequency range, which appeared relevant after studying the transfer function of the FOM. This requires apriori knowledge both about the FOM which is typically computationally hard to obtain and about the methodology used in our method to approximate that FOM. 
% In this work, we address one of the main issues of our direct optimization approach introduced in \cite{SchV2020}. There, we parametrize the system matrices of a low order pH system and then vary the parameters such that the transfer function of the parametrized system approximates the transfer function of the given FOM at a set sampling points $S:=\{s_i\}_{i=1\dots n_S} \subset \ri \R$, where $\ri \R$ denotes the imaginary axis. The purpose of approximating the given transfer function at $S$ is of course to obtain a good overall fit. However, the question on how to choose these sampling points was not addressed. Instead, in our numerical examples we have used a \emph{fixed set} of a large number (around 800) sampling points, logarithmically spaced throughout the frequency band which appeared relevant. The disadvantages of this approach are as follows.

Our article is organized as follows. In the next section, we formally introduce the terms needed to describe accuracy in the domain of MOR and summarize the main algorithm in \cite{SchV2020}. In Section~\ref{sec:ourmethod}, we explain how adaptive sampling is integrated into our method, explain the adaptive sampling algorithm and reformulate the new overall method. After that, we present a numerical study in Section~\ref{sec:numexperiments}, which investigates the behavior and highlights the effectiveness of the new adaptive sampling strategy. We outline future research directions in the context of the sampling based MOR method in Section~\ref{sec:conclusion}.

\section{Preliminaries}
\label{sec:background}

% Consider the function space
% \begin{align*} 
%   \mathcal{H}_\infty &:= \left\{ H : \C^+ \rightarrow {\mathbb C}^{p\times m} \; \bigg| \; H \text{ is analytic and } \right. \\ & \pushright{\left. \sup_{s \in \C^+} \left\| H(s) \right\|_2 < \infty \right\}},
% \end{align*}
In this work we consider the space $\cR\cH_\infty^{m \times m}$ of all real-rational and proper $m \times m$ transfer functions that have no poles in the set $\overline{\C^+} := \{\lambda \in \C \;|\; \Real(\lambda)\ge 0 \}$. For $H \in \cR\cH_\infty^{m \times m}$, the $\hinf$ norm is given by
\begin{align*}
  \left\| H \right\|_{\hinf} := \sup_{\lambda \in \C^{+}} \left\| H(\lambda) \right\|_{2} = \sup_{\omega \in \R}
  \sigma_{1}(H(\ri\omega)),
\end{align*}
where $\sigma_{1}(\cdot)$ denotes the largest singular value of its matrix valued argument. The $\hinf$ norm is often used to determine the approximation quality of a ROM due to the well-known error bound
\begin{align*}
  {\|y - \widetilde y \|}_{{L}^2([0,\infty),\R^m)} \le {\big\|H - \widetilde H\big\|}_{\hinf} \|u\|_{L^2([0,\infty),\R^m)},
\end{align*}
in which $y$ and $H$ and $\widetilde y$ and $\widetilde H$ denote the output and transfer function of a FOM and its ROM, respectively. Note that, alternatively, the $\mathcal{H}_2$ norm can be used as accuracy measure as it provides a similar error bound. Therefore, most MOR methods for linear systems aim at computing ROMs such that the error of the transfer functions with respect to these norms is small. In the following, we denote the Euclidian norm of the evaluated error transfer function by $E(s):=\big\|H(s)-\widetilde H(s)\big\|_2$. We focus our investigation on the $\hinf$ norm but the results for the $\mathcal{H}_2$ norm are similar, which can also be observed in \cite{SchV2020}.

While most MOR methods compute the ROM matrices via subspace projection, we construct a parameterized low order realization and treat MOR as a parameter optimization problem. A parameterized pH system is given as follows.
\begin{lem}[\cite{SchV2020}]
  Let $\theta \in \R^{n_\theta}$ be a vector with $n_\theta=\nx \left(\frac{3\nx+1}{2}+\enu \right)$. 
  Furthermore, let $\theta$ be partitioned as  $\theta:=\begin{bmatrix}\theta_J^\T,\,\theta_R^\T,\,\theta_Q^\T,\,\theta_B^\T\end{bmatrix}^\T$ with $\theta_J\in\R^{\nx(\nx-1)/2}$, $\theta_R\in\R^{\nx(\nx+1)/2},\,\theta_Q\in\R^{\nx(\nx+1)/2}$, and $\theta_B\in\R^{\nx \cdot \enu }$. Further define the matrices
  \begin{subequations}
  \begin{align}
    \pJ(\theta) &=\vtsu(\theta_J)^\T-\vtsu(\theta_J),\label{eq:pJ}\\
    \pR(\theta) &=\vtu(\theta_R)^\T  \vtu(\theta_R),\\
    \pQ(\theta) &=\vtu(\theta_Q)^\T  \vtu(\theta_Q),\label{eq:pQ}\\
    \pB(\theta) &=\vtf_{n,m}(\theta_B),
  \end{align}
  \end{subequations}
where the function $\vtu : \R^{\nx(\nx+1)/2} \rightarrow \R^{\nx \times \nx}$ maps a vector of length $\nx (\nx+1)/2$ to an upper triangular matrix, the function $\vtsu : \R^{\nx(\nx-1)/2} \rightarrow \R^{\nx \times \nx}$ maps a vector of length $\nx (\nx-1)/2$ to a strictly upper triangular matrix, and the function $\vtf_{n,m}: \R^{nm} \to \R^{n \times m}$ reshapes a vector of length $nm$ to an $n \times m$ matrix.
  Then, to each $\theta \in \R^{n_\theta}$ one can assign the pH system
  \begin{align} \label{eq:pHdelta}
    \pHsys(\theta): \;
    \begin{cases}
      \dot x(t) = \left( \pJ(\theta)-\pR(\theta) \right)\pQ(\theta) x(t)+ \pB(\theta) u(t),\\
      y(t) = \pB(\theta)^\T \pQ(\theta) x(t).
    \end{cases}
  \end{align}
 % with Hamiltonian
 % \begin{align}
 %   H(x(t), \theta) = \frac{1}{2}x(t)^\T Q(\theta) x(t).
 % \end{align}
  Conversely, to each pH system $\pHsys$ with $\nx$ states and $\enu $ inputs and outputs one can assign a vector $\theta \in \R^{n_\theta}$ such that $\pHsys = \pHsys(\theta)$ with $\pHsys(\theta)$ as in \eqref{eq:pHdelta}.
  %and its transfer function by
  %\begin{align}
  %  \pHtf(s,\theta)=\pB(\theta)^\T \pQ(\theta)\left(sI-\left(\pJ(\theta)-\pR(\theta)\right)\pQ(\theta)\right)^{-1}\pB(\theta),
  %\end{align}
  %or, equivalently,
  %\begin{align*}
  %  \pHtf(s,\theta)=\pB(\theta)^\T \left(s\pQ(\theta)^{-1}-\left(\pJ(\theta)-\pR(\theta)\right)\right)^{-1}\pB(\theta).
  %\end{align*}
\end{lem}
For details of the construction, we refer the reader to \cite{SchV2020}. In the following, we denote the transfer function of $\pHsys(\theta)$ by $\pHtf( \cdot, \theta)$.

The MOR method in \cite{SchV2020} consists of two stages. First, an initialization is performed to obtain the initial parameter vector $\theta_0$. Then the parameterization is optimized with respect to a sequence of objective functions designed to minimize the $\hinf$ norm of the difference between the transfer function of the given FOM and the transfer function of the parameterization.

The initialization step is based on a greedy interpolation strategy presented in \cite{beddig2019model}. In particular, the initial ROM is constructed by iteratively adding imaginary interpolation points at which the $\hinf$ norm of the current error transfer function is attained. This construction uses the projection introduced in \cite{Gugercin2012}. The $\hinf$ norm of the respective error transfer functions, which are large scale, are computed using the method proposed in \cite{SchV18}. Interpolation points are added, until the ROM has the state dimension chosen by the user. Then the parameter $\theta_0$ is extracted from this initial model to create the initial parameterized pH system and the optimization stage is started.

A direct gradient based minimization of the $\hinf$ error is avoided, since 
\begin{enumerate}
 \item the repeated $\hinf$ norm computation for large scale systems is computationally demanding despite efficient methods that were proposed recently,
 \item the $\hinf$ norm depends nonsmoothly on the model parameters, which obstructs the optimization, and
 \item the gradient of the $\hinf$ norm (if it exists) contains little information for a good descent direction because it only considers the error at a single point.
 \end{enumerate}
 Therefore, we propose to minimize instead the objective function
\begin{align}
  \begin{split}
    \pushleft{\loss\left(\gamma,H,\pHtf(\cdot, \theta),S\right) := }\\
    \pushright{\phantom{testtesttes}\frac{1}{\gamma}\sum\limits_{s_i\in S}\left(\left({\|H(s_i)-\pHtf(s_i,\theta)\|}_2-\gamma\right)_+\right)^2,}
    \label{eq:loss}
  \end{split}
\end{align}
where 
\begin{align*}
  ( \cdot )_+:  \R \rightarrow \overline{\R^+}, \quad x \mapsto 
  \begin{cases}
    x & \text{if } x\ge 0,\\
    0 & \text{if } x<0
  \end{cases}
\end{align*}
with respect to $\theta$ for decreasing values of $\gamma > 0$ to fit $\pHtf( \cdot, \theta)$ to a given transfer function $H$.
\begin{rem}
  A few remarks on $\loss$ are in order.
  \begin{enumerate}
    \item The value of $\loss$ is zero if the norm of the error between the given and the reduced transfer function is less than $\gamma$ at all sample points. In this way, a connection between $\loss$ and the $\hinf$ error minimization problem is established.
    \item Under mild assumptions it can be shown that $\loss$ is differentiable with respect to $\theta$ at some point $\theta_0~\in~\R^{n_\theta}$, which facilitates the optimization.
    \item The gradient of $\loss$ with respect to $\theta$ contains a descent direction that considers all sample points, at which the error is larger than the current level $\gamma$.
  \end{enumerate}
\end{rem}

The optimization step consists of first finding a suitable value for $\gamma$ and then minimizing $\loss$ using nonlinear optimization solvers (we use the implementation of BFGS from \cite{mogensen2018optim}). In this work, we select $\gamma$ using bisection, i.\,e.\ when the optimization determines a minimum at which $\loss$ is zero, we reduce $\gamma$ for the construction of the next optimization problem, otherwise it is increased. The method is similar to the method described in Alg.~\ref{alg:main}. However, Alg.~\ref{alg:main} includes the adaptive sampling strategy, which is explained in the following section.

\begin{figure*}[hbt]
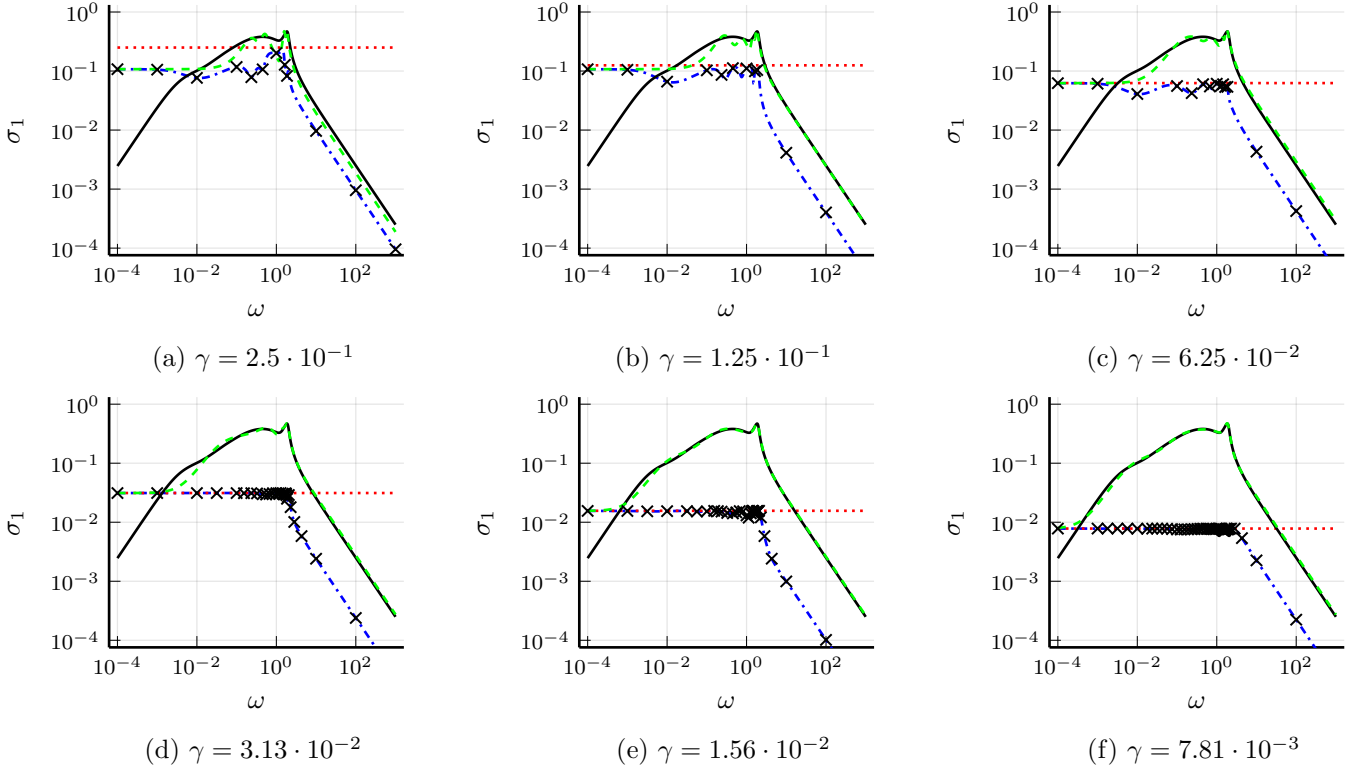

  \centering
  \begin{tabular}{ccc}
    \input{./PlotSources/IterationPlots/bisection8.1.tex} &
    \input{./PlotSources/IterationPlots/bisection8.2.tex} &
    \input{./PlotSources/IterationPlots/bisection8.3.tex} \\
    \phantom{test-set} (a) $\gamma = 2.5 \cdot 10^{-1}$ &\phantom{test-set}  (b) $\gamma = 1.25  \cdot 10^{-1}$& \phantom{test-set} (c) $\gamma = 6.25  \cdot 10^{-2}$ \\
    \input{./PlotSources/IterationPlots/bisection8.4.tex} &
    \input{./PlotSources/IterationPlots/bisection8.5.tex} &
    \input{./PlotSources/IterationPlots/bisection8.6.tex} \\
    \phantom{test-set} (d) $\gamma = 3.13 \cdot 10^{-2}$& \phantom{test-set} (e) $\gamma = 1.56 \cdot 10^{-2}$& \phantom{test-set} (f) $\gamma = 7.81 \cdot 10^{-3}$\\
  \end{tabular}
  \caption{The progress of Alg.~\ref{alg:main} for reduced model order of $r=8$ is depicted for decreasing levels $\gamma$. The FOM transfer function is illustrated as black solid line, the ROM transfer function is depicted as green dashed line, and the error is shown as blue dash-dotted line. The sample points are depicted as black crosses.}%
  \label{fig:adaptive_sampling}
\end{figure*}

\section{Our Method with Adaptive Sampling}
\label{sec:ourmethod}

Since a zero value of $\loss$ only guarantees that the error is less than $\gamma$ at the given sample points, the distribution of these points plays an essential role in connecting the minimization of $\loss$ to the minimization of the overall $\hinf$ error. If the sample points are distributed such that some peaks in the error transfer function are missed entirely, then a minimization of $\loss$ does not lead to a good $\hinf$ fit. This consideration may cause the user to prescribe a large number of sample points. However, the cost of evaluating $\loss$ and its gradient grows with the number of sample points, so a large number of unnecessary sample points should also be avoided. For these reasons, an automatic selection of the sample points is desirable.

We propose to construct the automatic sample selection algorithm based on the following result from \cite{Apk2018}. Therein, for $\omega_1 \le \omega_2$, $L[\omega_1,\omega_2]$ is called a first-order upper bound for a continuously differentiable $\phi: \R  \rightarrow \overline{\R^+}$, if $L[\omega_1,\omega_2] \ge |\phi'(\omega)|$ for all $\omega \in [\omega_1,\omega_2]$.
\begin{lem}[\cite{Apk2018}] 
  Let $\phi: \R  \rightarrow \overline{\R^+}$ be continuously differentiable and let $\omega_i$ and $\omega_{i+1}$ be two consecutive nodes of a piecewise linear interpolation $P_\phi$ of $\phi$ such that $\gamma^* \ge \max \{\phi(\omega_i), \phi(\omega_{i+1})\}$. Furthermore, let  $L[ \omega_i,  \omega_{i+1}]$ be a first-order upper bound for $\phi$ on the interval $[ \omega_i, \omega_{i+1}]$. Assume that for a given tolerance $\tau > 0$ we have that
    \begin{align*}
      L[ \omega_i, \omega_{i+1} ] (\omega_{i+1} - \omega_i) < 2 \gamma^* + 2 \tau - \phi(\omega_i) - \phi(\omega_{i+1}).
    \end{align*}
    Then, $\phi(\omega) < \gamma^*+\tau$ for every $\omega \in [\omega_i, \omega_{i+1}]$.
    \label{lem:accuracy}
\end{lem}
In \cite{Apk2018} this result is used to construct an accuracy certificate after their main optimization algorithm (which computes $\hinf$ controllers) has terminated to restart it on a finer grid, if necessary. Since our method, on the other hand, is based on multiple optimizations of $\loss$ for varying values of $\gamma$, and, moreover, a meaningful tolerance is also given by the target value for $\gamma$, we can use Lemma \ref{lem:accuracy} already within our algorithm to update the sample points as $\gamma$ is decreased before the nonlinear optimization step is started.

\begin{algorithm}[b]
  \caption{Logarithmic Sampling Adaptation}
  \label{alg:samp}
  \begin{algorithmic}[1]
    \REQUIRE{The error $E : \overline{\C^+}  \rightarrow \overline{\R^+}$, imaginary parts of current sample points $S = \{\omega_1,\,\ldots,\,\omega_{n_S}\}$, current level $\gamma>0$.}
    \ENSURE{Updated set of sample points $S$.}
    \STATE{Set $n_{\text{new}}:=1$.}
    \WHILE{$n_{\text{new}}>0$}
      \STATE{Sort $\{\omega_1,\,\ldots,\,\omega_{n_S}\}$ such that $\omega_i<\omega_{i+1}$ for all $i\in 1, \dots, n_S$.}
      \STATE{Set $n_{\text{new}}:=0$.}
      \FOR{$i = 1,\, \dots,\, n_S-1$}
        \STATE{Set $\omega_{\text{test}}:=10^{(\log_{10}(\omega_i)+\log_{10}(\omega_{i+1}))/2}$.}
        \STATE{Set $d_1 := (E(\ri \omega_{\text{test}})-E(\ri \omega_i))/(\omega_{\text{test}}-\omega_i)$.}
        \STATE{Set $d_2 := (E(\ri \omega_{i+1})-E(\ri \omega_{\text{test}}))/(\omega_{i+1}-\omega_{\text{test}})$.}
        \STATE{Set $\gamma^* := \max\{E(\ri \omega_{i}), E(\ri \omega_{i+1})\}$.}
        \STATE{Set $d^* := \max\{d_1, d_2\}$.}
        \IF{$d^* (\omega_{i+1}-\omega_i) \ge 2(\gamma^* + \gamma)-(E(\ri \omega_i)+E(\ri \omega_{i+1}))$}
          \STATE{Set $n_{\text{new}}:=n_{\text{new}}+1$.}
          \STATE{Set $n_S:=n_S+1$ and set $S:= S \cup \{\omega_{\text{test}}\}$.}
        \ENDIF
      \ENDFOR
    \ENDWHILE
  \end{algorithmic}
\end{algorithm}

We use Alg.~\ref{alg:samp} to update the sample points that are used in the objective in Alg.~\ref{alg:main} to achieve a good $\hinf$ fit. The sampling update in Alg.~\ref{alg:samp} works as follows. Between each pair of adjacent sample points, a new sample $\ri \omega_{\text{test}}$ is chosen logarithmically and the difference quotients between this new sample and the two respective adjacent points are computed. The maximum of these difference quotients serves as estimate for the first-order upper bound in the respective interval. The tolerance ($\tau$ in Lemma~\ref{lem:accuracy}) is set to the current level $\gamma$. If now the criterion given in the inequality in \eqref{lem:accuracy} which guarantees that the error between the two sample points is less than $\gamma^*+\tau$ is not met, the extra sample point $\ri \omega_{\text{test}}$ is added to the sample set. Due to the fact that multiple additional points may be required between two original sample points, we iterate through the set of sample points until no more new points are added. Note that in Alg.~\ref{alg:samp}, for simplicity, we only refer to the imaginary parts of the sample points.

\begin{rem}
  \cite{Apk2018} use the difference quotients between the two adjacent sample points already in the set as estimate for the derivative upper bound. However, since our method tends to cause the error function to attain similar values (in magnitude) at the sample points, the actual derivative is heavily underestimated, when this difference quotient is used. Inserting the extra point between the two samples, on the other hand, provides better estimates for the derivative.
\end{rem}

\begin{algorithm}[bt]
  \caption{Iterative Reduction Algorithm}
  \label{alg:main}
  \begin{algorithmic}[1]
    \REQUIRE{Transfer function $H \in \cR\cH_\infty^{m \times m}$, initial parameterized transfer function $\pHtf( \cdot, \theta)\in \cR\cH_\infty^{m \times m}$, $\gamma_{\max}>0$, initial sample points $S$, bisection tolerance $\tau_{\rm b}$.}
    \ENSURE{Optimized ROM transfer function $\pHtf( \cdot, \theta_{\rm opt})$}
    \STATE{Set $\gamma_{\min}:=0.$}
    \WHILE{$\gamma \tau_{\rm b} < (\gamma_{\max}-\gamma_{\min})$}
      \STATE{Set $\gamma:=(\gamma_{\max}-\gamma_{\min}))/2$.}
      \STATE{Update $S$ using Alg.~\ref{alg:samp}.}
      \STATE{Update $\theta_{\rm opt} := \argmin\limits_{\theta \in \R^{n_\theta}} \loss(\gamma, H, \pHtf( \cdot, \theta), S)$.}
      \IF{$\loss(\gamma, H, \pHtf( \cdot, \theta_{\rm opt}), S) > 0$.}
        \STATE{Set $\gamma_{\min}:=\gamma$.}
      \ELSE
        \STATE{Set $\gamma_{\max}:=\gamma$.}
      \ENDIF
    \ENDWHILE
  \end{algorithmic}
\end{algorithm}

In Alg.~\ref{alg:main}, our adaptive sampling MOR algorithm based on a bisection over $\gamma$ is described. In each iteration, first the sample points are updated as described above. Then a minimizer $\theta_{\rm opt}$ is computed for the objective function for the current $\gamma$ level. For that, we use a BFGS optimization solver from \cite{mogensen2018optim}. If the final objective value is zero, i.\,e., the error at all sample points is below $\gamma$, then $\gamma_{\max}$ is updated to the current value of $\gamma$ such that $\gamma$ is reduced in the next iteration. Otherwise, the currently required accuracy is not met and $\gamma$ is increased by setting $\gamma_{\min}$ to the current value of $\gamma$.

\section{Numerical Experiments}
\label{sec:numexperiments}

We evaluate our new adaptive method on a pH mass-springer-damper chain benchmark example also used in \cite{Gugercin2012} to assess the performance of the pH structure preserving variant of IRKA. We set the state dimension only to 100 to be able to compute the global $\hinf$ approximation error during the evaluation. However, we demonstrate in our previous work (\cite{SchV2020}) that for larger state dimensions, neither the shape of the transfer function of the given pH model nor the performance of our reduction algorithm change significantly. The input and output dimension of the model is 2. The purpose of our experiments is (1) to demonstrate that the $\hinf$ accuracy of the adaptive method is comparable to our previous method despite using significantly fewer sample points and (2) to investigate where the adaptive sampling method places the sample points.

Our experimental setup is as follows. We compute ROMs for the pH benchmark system at reduced state dimensions $4,\, 6, \,\dots,\, 20$. For that, we run our greedy interpolation initialization procedure and stop at $2,\, 3,\, \dots,\, 10$ interpolation points. After that, the initial parameter $\theta_0$ is extracted from the initial reduced model and the initial parameterized model is constructed. As initial set of sample points, we concatenate a set of logarithmically spaced points $\{10^{-8}, 10^{-7}, \dots, 10^{5}\}$ with the interpolation points computed during initialization. For the bisection parameters we choose $\tau_{\rm b}=0.1$ and initially set $\gamma_{\max}=0.5$. Note that $\gamma_{\min}$ is always set to zero. All experiments are run in \texttt{julia} (version 1.5.3) on an Intel\textsuperscript{\textregistered}\,Core\texttrademark\; i9-9900K CPU at 3.60\,GHz with 32\,GB of RAM.

% Note that each interpolation point increases the reduced state dimension by two, since we only require an interpolation property in the direction of the maximum singular vectors but must split the interpolation subspace into its real and imaginary part.

In Fig.~\ref{fig:adaptive_sampling} we illustrate the progress of our method over the first 8 iterations, as $\gamma$ is decreasing. At early stages of the algorithm (Fig.~\ref{fig:adaptive_sampling} (a-c)), the error at the different sample points varies, while it attains similar values close to $\gamma$ at later iterations, except beyond a certain frequency at which all depicted transfer functions converge to zero. This rather flat error curve indicates that a good $\hinf$ fit is found. Furthermore, most sample points are added around $10^0$, which is where the transfer function of the FOM has two peaks.
%We further investigate the distribution of the selected sample points in Fig.~\ref{fig:adaptive_sampling}.

\begin{figure}[tb]
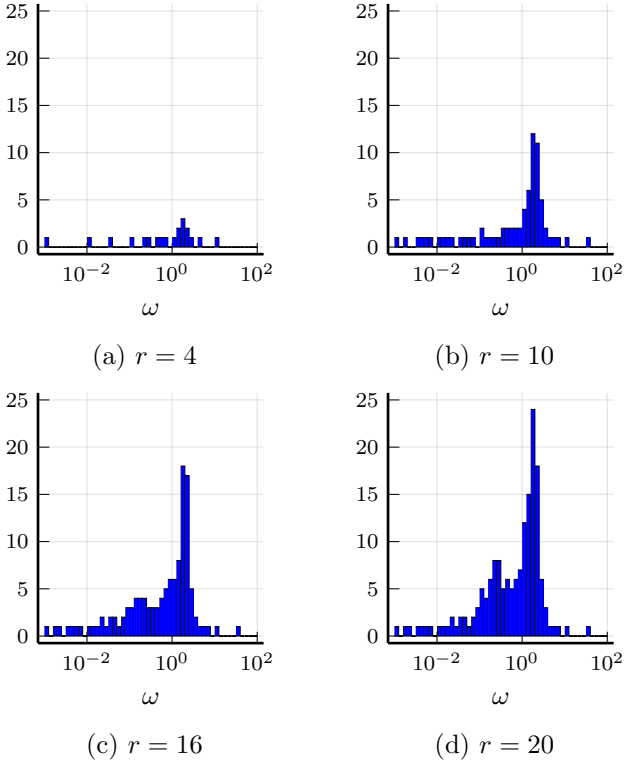

  \centering
  \begin{tabular}{ccc}
    \input{./PlotSources/SampleHistograms/samplehistogram4.tex} &
    \input{./PlotSources/SampleHistograms/samplehistogram10.tex} \\
    (a) $r=4$ & (b) $r=10$ \\
    \input{./PlotSources/SampleHistograms/samplehistogram16.tex} &
    \input{./PlotSources/SampleHistograms/samplehistogram20.tex} \\
    (c) $r=16$ & (d) $r=20$ 
  \end{tabular}
  \caption{Histograms showing the distribution of the sample points after Alg.~\ref{alg:main} has terminated for different reduced model orders.}%
  \label{fig:histogram}
\end{figure}

\begin{figure}[tb]
  \centering
  \input{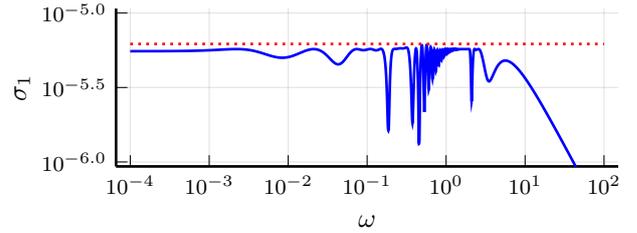}
  \caption{Final error transfer function for reduced model order $r=20$ depicted in blue. The corresponding $\gamma$ level is shown as red dotted line.}%
  \label{fig:r20errors}
\end{figure}

In Fig.~\ref{fig:histogram} histograms showing the distribution of the sample points after the completion of Alg.~\ref{alg:main} are depicted for different reduced model orders and, consequently, different final accuracies. % Note that the total numbers of sample points are given in Tab.~\ref{tab:num_samples}.
We can observe that as the reduced model order and at the same time the final accuracy increases, the concentration of sample points around $10^0$ intensifies. This is due to the fact that the error has many small peaks in this frequency range, which lead to a higher estimate of the first-order derivative bound. This can also be observed in Fig.~\ref{fig:r20errors}, in which the spiky behavior of the error transfer function between $10^{-1}$ and $10^1$ is illustrated.
% We conclude from this concentration that the shape of the given transfer function around $10^0$ is particularly hard to approximate with a low system dimension.

\begin{table}[tb]
  \centering
    \caption{Number of sample points for different reduced model orders during the execution of Alg.~\ref{alg:main} at different intermediate values for $\gamma$.}
  \label{tab:num_samples}
  \begin{tabular}{c||c|c|c|c|c|c|c|c|c}
    \diagbox{$\gamma$}{$r$} & 4  & 6  & 8  & 10 & 12 & 14  & 16  & 18  & 20 \\ \hline
 2.50e-01               & 16 & 17 & 18 & 19 & 20 & 21  & 22  & 23  & 24  \\
 1.25e-01               & 16 & 17 & 18 & 19 & 20 & 21  & 22  & 23  & 24  \\
 6.25e-02               & 21 & 21 & 21 & 22 & 22 & 23  & 23  & 23  & 25  \\
 3.13e-02               & -  & 38 & 36 & 34 & 29 & 30  & 27  & 24  & 26  \\
 1.56e-02               & -  & -  & 44 & 46 & 54 & 53  & 53  & 45  & 32  \\
 7.81e-03               & -  & -  & 57 & 59 & 63 & 70  & 75  & 65  & 68  \\
 3.91e-03               & -  & -  & 63 & 63 & 66 & 74  & 81  & 78  & 106  \\
 1.95e-03               & -  & -  & -  & 66 & 67 & 78  & 87  & 84  & 108  \\
 9.77e-04               & -  & -  & -  & 76 & 76 & 83  & 94  & 88  & 111  \\
 4.88e-04               & -  & -  & -  & 81 & 82 & 90  & 95  & 97  & 118  \\
 2.44e-04               & -  & -  & -  & -  & 88 & 91  & 99  & 101 & 121  \\
 1.22e-04               & -  & -  & -  & -  & 94 & 99  & 104 & 103 & 123  \\
 6.10e-05               & -  & -  & -  & -  & -  & 103 & 110 & 106 & 128  \\
 3.05e-05               & -  & -  & -  & -  & -  & 110 & 115 & 113 & 133  \\
 1.53e-05               & -  & -  & -  & -  & -  & -   & 128 & 124 & 144  \\
 7.63e-06               & -  & -  & -  & -  & -  & -   & -   & 132 & 152  \\
 3.81e-06               & -  & -  & -  & -  & -  & -   & -   & -   & 159  \\

  \end{tabular}
\end{table}

In Tab.~\ref{tab:num_samples}, the number of samples for each model order and value for $\gamma$ that is attained during optimization is given. Note that we only show the $\gamma$ values as long as the bisection leads to a decreasing value, since no points are deleted, when $\gamma$ is increased. The increased number of sample points for larger reduced model orders in the first row is due to the initialization, which can use one more interpolation point per state dimension. It is remarkable that despite the fact that $\gamma$ decreases over several orders of magnitude, the number of sample points stays moderate and typically only a few sample points are added as $\gamma$ is halved.

This only slight increase in the number of samples is due to our iterative decrease in $\gamma$. Since the accuracy of the ROM is gradually increased, the criterion in line 11 of Alg.~\ref{alg:samp}, requires fewer sample points even for a smaller tolerance (which is always set to $\gamma$). If we, instead, were to fix $\gamma$ to the final accuracy and consult Alg.~\ref{alg:samp} for the sample points based on the initial model, the number of required sample points is drastically higher, from $38$ samples for $r=4$, via $8\,962$ for $r=12$ to $160\,536$ samples for $r=20$. Such a large number of sample points would make the evaluation of $\loss$ and thus its gradient based minimization prohibitively expensive. This underlines the importance of combining the adaptive sampling strategy with our iterative reduction of $\gamma$ in our objective function $\loss$.

\begin{table}[tb]
  \centering
  \caption{Runtime comparison between the adaptive sampling based Alg.~\ref{alg:main} and its fixed sampling counterpart.}
  \label{tab:runtimes}
  \begin{tabular}{c||c|c|c}
    & \multicolumn{2}{l}{runtime time in seconds} & \\
    $r$ & fixed sampling & Alg.~\ref{alg:main} & ratio \\ \hline
     % 4 & 4.50e+01$\,\mathrm{s}$ & 2.40e+00$\,\mathrm{s}$ & 1.87e+01 \\ 
 % 6 & 1.61e+02$\,\mathrm{s}$ & 8.44e+00$\,\mathrm{s}$ & 1.91e+01 \\ 
 % 8 & 1.66e+02$\,\mathrm{s}$ & 6.06e+01$\,\mathrm{s}$ & 2.73e+00 \\ 
% 10 & 9.26e+02$\,\mathrm{s}$ & 1.29e+02$\,\mathrm{s}$ & 7.16e+00 \\ 
% 12 & 9.46e+02$\,\mathrm{s}$ & 1.94e+02$\,\mathrm{s}$ & 4.87e+00 \\ 
% 14 & 1.13e+03$\,\mathrm{s}$ & 4.76e+02$\,\mathrm{s}$ & 2.37e+00 \\ 
% 16 & 4.21e+03$\,\mathrm{s}$ & 3.09e+02$\,\mathrm{s}$ & 1.36e+01 \\ 
% 18 & 3.64e+03$\,\mathrm{s}$ & 1.10e+03$\,\mathrm{s}$ & 3.32e+00 \\ 
% 20 & 5.39e+03$\,\mathrm{s}$ & 1.58e+03$\,\mathrm{s}$ & 3.40e+00 \\ 
 4 & 4.50e+01 & 2.40e+00 & 1.87e+01 \\ 
 6 & 1.61e+02 & 8.44e+00 & 1.91e+01 \\ 
 8 & 1.66e+02 & 6.06e+01 & 2.73e+00 \\ 
10 & 9.26e+02 & 1.29e+02 & 7.16e+00 \\ 
12 & 9.46e+02 & 1.94e+02 & 4.87e+00 \\ 
14 & 1.13e+03 & 4.76e+02 & 2.37e+00 \\ 
16 & 4.21e+03 & 3.09e+02 & 1.36e+01 \\ 
18 & 3.64e+03 & 1.10e+03 & 3.32e+00 \\ 
20 & 5.39e+03 & 1.58e+03 & 3.40e+00 \\ 

  \end{tabular}
\end{table}

\begin{figure}[tb]
  \centering
  \begin{tikzpicture}[/tikz/background rectangle/.style={fill={rgb,1:red,1.0;green,1.0;blue,1.0}, draw opacity={1.0}}, show background rectangle]
\begin{axis}[title={}, title style={at={{(0.5,1)}}, font={{\fontsize{14 pt}{18.2 pt}\selectfont}}, color={rgb,1:red,0.0;green,0.0;blue,0.0}, draw opacity={1.0}, rotate={0.0}}, legend columns=2, legend style={color={rgb,1:red,0.0;green,0.0;blue,0.0}, draw opacity={1.0}, line width={1}, solid, fill={rgb,1:red,1.0;green,1.0;blue,1.0}, fill opacity={1.0}, text opacity={1.0}, font={{\fontsize{8 pt}{10.4 pt}\selectfont}}, at={(1, 1.02)}, anchor={south east}}, axis background/.style={fill={rgb,1:red,1.0;green,1.0;blue,1.0}, opacity={1.0}}, anchor={north west}, xshift={1.0mm}, yshift={-1.0mm}, width={0.45\textwidth}, height={0.3\textheight}, scaled x ticks={false}, xlabel={$r$}, x tick style={color={rgb,1:red,0.0;green,0.0;blue,0.0}, opacity={1.0}}, x tick label style={color={rgb,1:red,0.0;green,0.0;blue,0.0}, opacity={1.0}, rotate={0}}, xlabel style={}, xmajorgrids={true}, xmin={3.52}, xmax={20.48}, xtick={{4.0,8.0,12.0,16.0,20.0}}, xticklabels={{$4$,$8$,$12$,$16$,$20$}}, xtick align={inside}, xticklabel style={font={{\fontsize{8 pt}{10.4 pt}\selectfont}}, color={rgb,1:red,0.0;green,0.0;blue,0.0}, draw opacity={1.0}, rotate={0.0}}, x grid style={color={rgb,1:red,0.0;green,0.0;blue,0.0}, draw opacity={0.1}, line width={0.5}, solid}, axis x line*={left}, x axis line style={color={rgb,1:red,0.0;green,0.0;blue,0.0}, draw opacity={1.0}, line width={1}, solid}, scaled y ticks={false}, ylabel={$\mathcal{H}_\infty$ error}, y tick style={color={rgb,1:red,0.0;green,0.0;blue,0.0}, opacity={1.0}}, y tick label style={color={rgb,1:red,0.0;green,0.0;blue,0.0}, opacity={1.0}, rotate={0}}, ylabel style={}, ymode={log}, log basis y={10}, ymajorgrids={true}, ymin={4.169570177021927e-6}, ymax={0.7106243024441214}, ytick={{1.0e-5,0.0001,0.001,0.01,0.1}}, yticklabels={{$10^{-5}$,$10^{-4}$,$10^{-3}$,$10^{-2}$,$10^{-1}$}}, ytick align={inside}, yticklabel style={font={{\fontsize{8 pt}{10.4 pt}\selectfont}}, color={rgb,1:red,0.0;green,0.0;blue,0.0}, draw opacity={1.0}, rotate={0.0}}, y grid style={color={rgb,1:red,0.0;green,0.0;blue,0.0}, draw opacity={0.1}, line width={0.5}, solid}, axis y line*={left}, y axis line style={color={rgb,1:red,0.0;green,0.0;blue,0.0}, draw opacity={1.0}, line width={1}, solid}, colorbar style={title={}}, point meta max={nan}, point meta min={nan}]
    \addplot[color=black, name path={1c97c6ec-a4e8-4e04-b96d-a14d666e4a38}, draw opacity={1.0}, line width={1}, solid, mark={square}, mark size={3.0 pt}, mark options={color={rgb,1:red,0.0;green,0.0;blue,0.0}, draw opacity={1.0}, fill=black, fill opacity={1.0}, line width={0.75}, rotate={0}, solid}]
        coordinates {
            (4,0.15753775527276784)
            (6,0.04027322931391762)
            (8,0.009039266088716613)
            (10,0.0013965612409578953)
            (12,0.00036038074920490167)
            (14,9.760860161700115e-5)
            (16,3.718258464869135e-5)
            (18,1.4904130437195738e-5)
            (20,9.515525824944514e-6)
        }
        ;
    \addlegendentry {BT}
    \addplot[color=blue, name path={fdc86282-b401-4979-abae-4468f1aaf760}, draw opacity={1.0}, line width={1}, solid, mark={square*}, mark size={3.0 pt}, mark options={color={rgb,1:red,0.0;green,0.0;blue,0.0}, draw opacity={1.0}, fill=blue, fill opacity={1.0}, line width={0.75}, rotate={0}, solid}]
        coordinates {
            (4,0.4813471942945101)
            (6,0.5053326346940811)
            (8,0.33265402950983736)
            (10,0.2792712148058059)
            (12,0.2535149168194241)
            (14,0.1870396561695439)
            (16,0.14977502520076516)
            (18,0.12700565585039805)
            (20,0.086946101277349)
        }
        ;
    \addlegendentry {pH-BT}
    \addplot[color=gray, name path={8bcb0643-48fc-42f0-bc69-1dd8a3c976d3}, draw opacity={1.0}, line width={1}, solid, mark={diamond}, mark size={3.0 pt}, mark options={color={rgb,1:red,0.0;green,0.0;blue,0.0}, draw opacity={1.0}, fill=gray, fill opacity={1.0}, line width={0.75}, rotate={0}, solid}]
        coordinates {
            (4,0.17027228081798673)
            (6,0.12484574367356895)
            (8,0.04303408150106201)
            (10,0.0018477730009792128)
            (12,0.0025173078149703)
            (14,0.00011531401245745243)
            (16,0.0003000306047994916)
            (18,2.2769436542425816e-5)
            (20,1.0259198015791216e-5)
        }
        ;
    \addlegendentry {IRKA}
    \addplot[color=orange, name path={5e7e1fa2-9a18-4027-b9a8-5d99613a1c14}, draw opacity={1.0}, line width={1}, solid, mark={diamond*}, mark size={3.0 pt}, mark options={color={rgb,1:red,0.0;green,0.0;blue,0.0}, draw opacity={1.0}, fill=orange, fill opacity={1.0}, line width={0.75}, rotate={0}, solid}]
        coordinates {
            (4,0.29644543479395774)
            (6,0.2520497943209842)
            (8,0.16276581594136239)
            (10,0.1287240783673755)
            (12,0.09709961195728065)
            (14,0.07263289970224365)
            (16,0.05774756684325709)
            (18,0.04021116428444131)
            (20,0.027135580825011017)
        }
        ;
    \addlegendentry {pH-IRKA}
    \addplot[color=red, name path={ee35c385-852e-4ea4-bc93-4a19681d6410}, draw opacity={1.0}, line width={1}, solid, mark={triangle}, mark size={3.0 pt}, mark options={color={rgb,1:red,0.0;green,0.0;blue,0.0}, draw opacity={1.0}, fill=red, fill opacity={1.0}, line width={0.75}, rotate={0}, solid}]
        coordinates {
            (4,0.1027537666664209)
            (6,0.05002706591150752)
            (8,0.0063298755220841055)
            (10,0.0009332812813081077)
            (12,0.0002534960529786643)
            (14,7.527200720200445e-5)
            (16,2.5456121200930722e-5)
            (18,8.969214861824735e-6)
            (20,5.8634602962126474e-6)
        }
        ;
      \addlegendentry {prev. paper}
    \addplot[color=green, name path={ee35c385-852e-4ea4-bc93-4a19681d6410}, draw opacity={1.0}, line width={1}, solid, mark={triangle*}, mark size={3.0 pt}, mark options={color={rgb,1:red,0.0;green,0.0;blue,0.0}, draw opacity={1.0}, fill=green, fill opacity={1.0}, line width={0.75}, rotate={0}, solid}]
        coordinates {
        (4, 0.07699050483017127)
        (6, 0.034230254544311854)
        (8, 0.005515885865706487)
        (10, 0.0006775736340773168)
        (12, 0.00016893491608513286)
        (14, 5.0888316561885745e-5)
        (16, 1.605229213459533e-5)
        (18, 1.1024396483311035e-5)
        (20, 6.093491399028249e-6)
        }
        ;
    \addlegendentry {our method}
\end{axis}
\end{tikzpicture}
  \caption{$\mathcal{H}_\infty $ error comparison}
  \label{fig:hinferrors}
\end{figure}
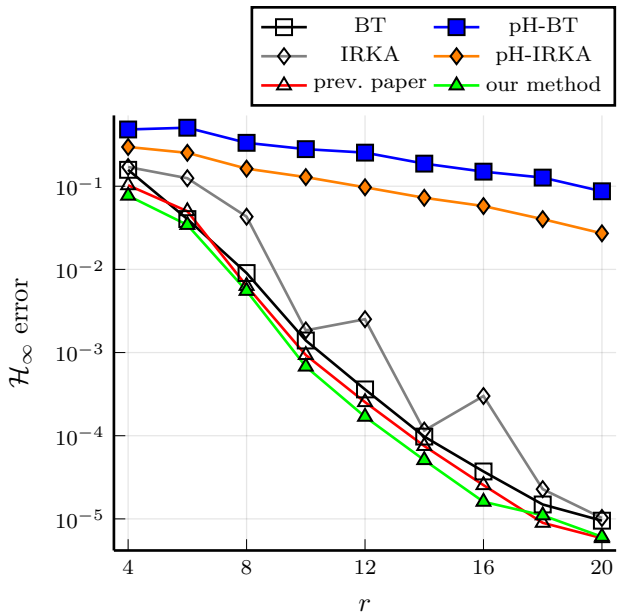

In Tab.~\ref{tab:runtimes}, the runtimes of Alg.~\ref{alg:main} for different reduced model orders are reported and compared to a fixed sampling variant of this algorithm, in which around 800 logarithmically spaced sample points are chosen before the optimization is started. The speedup ranges from around $3\times$ to $19\times$ and we note that for the two smallest reduced model orders the speedup is the largest. For each reduced order except for $r=18$ the number of optimizations (line 5 in Alg.~\ref{alg:main}) was the same for the fixed sampling and the adaptive sampling approach and also the resulting $\hinf$ errors, which are studied in Fig.~\ref{fig:hinferrors}, are approximately the same.

Fig.~\ref{fig:hinferrors} emphasizes that our new adaptive method leads to a similar accuracy compared to our previous method with a large fixed number of samples. We also mention that our method yields structure preserving ROMs with an orders of magnitude higher accuracy compared to the other structure preserving MOR algorithms pH-BT and pH-IRKA and even outperforms the non-structure preserving methods BT and IRKA for all reduced model orders considered in this comparison.
% However, the new spectral-factorization based methods by \cite{BreitenU2021} improve the accuracy of structure preserving MOR algorithms significantly. In their current preprint they report an accuracy comparable to the accuracy of IRKA.

\section{Conclusion}
\label{sec:conclusion}

We have equipped our optimization-based structure preserving MOR algorithm from \cite{SchV2020} with an adaptive sampling strategy to increase its usability and lower its computational footprint. In our numerical experiments, we have demonstrated that combining this adaptive sampling with our reduction algorithm, which is based on an objective function that iteratively increases the overall accuracy, is beneficial to keep the number of samples low while retaining a high accuracy.

Future research includes extending the method to parameter dependent systems as well as studying additional model structures. Finally, the improved accuracy (also compared to the unstructured MOR algorithms) motivates studying the accuracy of our method also for unstructured systems.

\bibliography{ifacconf}

\begin{thebibliography}{12}
\providecommand{\natexlab}[1]{#1}
\providecommand{\url}[1]{\texttt{#1}}
\providecommand{\urlprefix}{URL }
\expandafter\ifx\csname urlstyle\endcsname\relax
  \providecommand{\doi}[1]{doi:\discretionary{}{}{}#1}\else
  \providecommand{\doi}{doi:\discretionary{}{}{}\begingroup
  \urlstyle{rm}\Url}\fi

\bibitem[{Antoulas(2005)}]{Ant05}
Antoulas, A.C. (2005).
\newblock \emph{Approximation of Large-Scale Dynamical Systems}, volume~6 of
  \emph{Adv. Des. Control}.
\newblock SIAM Publications, Philadelphia, PA.

\bibitem[{Apkarian and Noll(2018)}]{Apk2018}
Apkarian, P. and Noll, D. (2018).
\newblock Structured $\mathcal{H}_\infty$-control of infinite-dimensional
  systems.
\newblock \emph{Internat. J. Robust Nonlinear Control}, 28(9), 3212--3238.

\bibitem[{Beattie et~al.(2019)Beattie, Mehrmann, and {Van
  Dooren}}]{Beattie2019}
Beattie, C., Mehrmann, V., and {Van Dooren}, P. (2019).
\newblock Robust port-{H}amiltonian representations of passive systems.
\newblock \emph{Automatica J. IFAC}, 100, 182--186.

\bibitem[{Beddig et~al.(2019)Beddig, Benner, Dorschky, Reis, Schwerdtner,
  Voigt, and Werner}]{beddig2019model}
Beddig, R.S., Benner, P., Dorschky, I., Reis, T., Schwerdtner, P., Voigt, M.,
  and Werner, S.W. (2019).
\newblock Model reduction for second-order dynamical systems revisited.
\newblock \emph{PAMM Proc. Appl. Math. Mech.}, 19(1), e201900224.

\bibitem[{Breiten and Unger(2021)}]{BreitenU2021}
Breiten, T. and Unger, B. (2021).
\newblock Passivity preserving model reduction via spectral factorization.
\newblock Preprint arXiv:2103.13194.

\bibitem[{Gugercin et~al.(2012)Gugercin, Polyuga, Beattie, and van~der
  Schaft}]{Gugercin2012}
Gugercin, S., Polyuga, R.V., Beattie, C., and van~der Schaft, A. (2012).
\newblock Structure-preserving tangential interpolation for model reduction of
  port-{Hamiltonian} systems.
\newblock \emph{Automatica J. IFAC}, 48(9), 1963--1974.

\bibitem[{Mogensen and Riseth(2018)}]{mogensen2018optim}
Mogensen, P.K. and Riseth, A.N. (2018).
\newblock Optim: A mathematical optimization package for {Julia}.
\newblock \emph{J. Open Source Softw.}, 3(24), 615--618.

\bibitem[{Polyuga and van~der Schaft(2008)}]{Polyuga2008}
Polyuga, R.V. and van~der Schaft, A.J. (2008).
\newblock Structure preserving model reduction of port-{H}amiltonian systems.
\newblock In \emph{Proceedings of the 18th International Symposium on
  Mathematical Theory of Networks and Systems}. Blacksburg, VA.

\bibitem[{Schwerdtner and Voigt(2018)}]{SchV18}
Schwerdtner, P. and Voigt, M. (2018).
\newblock Computation of the $\mathcal{L}_\infty$-norm using rational
  interpolation.
\newblock \emph{IFAC-PapersOnLine}, 51(25), 84--89.
\newblock 9th IFAC Symposium on Robust Control Design 2018, Florian\'opolis,
  Brazil.

\bibitem[{Schwerdtner and Voigt(2020)}]{SchV2020}
Schwerdtner, P. and Voigt, M. (2020).
\newblock Structure preserving model order reduction by parameter optimization.
\newblock Preprint arXiv:2011.07567.

\bibitem[{Serhani et~al.(2019)Serhani, Matignon, and Haine}]{Serhani2019}
Serhani, A., Matignon, D., and Haine, G. (2019).
\newblock A partitioned finite element method for the structure-preserving
  discretization of damped infinite-dimensional port-{H}amiltonian systems with
  boundary control.
\newblock In F.~Nielsen and F.~Barbaresco (eds.), \emph{Geometric Science of
  Information}, 549--558. Springer International Publishing, Cham.

\bibitem[{van~der Schaft and Jeltsema(2014)}]{Schaft2014}
van~der Schaft, A. and Jeltsema, D. (2014).
\newblock Port-{H}amiltonian systems theory: An introductory overview.
\newblock \emph{Found. Trends Systems Control}, 1(2--3), 173--378.

\end{thebibliography}

\end{document}